\documentclass[twocolumn,secnumarabic,amssymb, nobibnotes, aps, prd,floatfix]{revtex4-2}

\usepackage{amsmath}
\usepackage{amssymb}
\usepackage{todonotes}
\usepackage{hyperref}

\begin{document}

\title{The impact of kinetic inductance on the critical current oscillations \\ of nanobridge SQUIDs}%

\author{H. Dausy}
\email{heleen.dausy@kuleuven.be}
\affiliation{Quantum Solid State Physics, Department of Physics and Astronomy, KU Leuven, Celestijnenlaan 200D, B-3001 Leuven, Belgium}
\author{L. Nulens}
\affiliation{Quantum Solid State Physics, Department of Physics and Astronomy, KU Leuven, Celestijnenlaan 200D, B-3001 Leuven, Belgium}
\author{B. Raes}
\affiliation{Quantum Solid State Physics, Department of Physics and Astronomy, KU Leuven, Celestijnenlaan 200D, B-3001 Leuven, Belgium}
\author{M.J. Van Bael}
\affiliation{Quantum Solid State Physics, Department of Physics and Astronomy, KU Leuven, Celestijnenlaan 200D, B-3001 Leuven, Belgium}
\author{J. Van de Vondel}
\affiliation{Quantum Solid State Physics, Department of Physics and Astronomy, KU Leuven, Celestijnenlaan 200D, B-3001 Leuven, Belgium}

\begin{abstract}
In this work, we study the current phase relation (C$\Phi$R) of lithographically fabricated molybdenum germanium (Mo$_{79}$Ge$_{21}$) nanobridges, which is intimately linked to the nanobridge kinetic inductance. We do this by imbedding the nanobridges in a SQUID. We observe that for temperatures far below $T_c$, the C$\Phi$R is linear as long as the condensate is not weakened by the presence of supercurrent. We demonstrate lithographic control over the nanobridge kinetic inductance, which scales with the nanobridge aspect ratio. This allows to tune the SQUID $I_c(B)$ characteristic. The SQUID properties that can be controlled in this way include the SQUID sensitivity and the positions of the critical current maxima. These observations can be of use for the design and operation of future superconducting devices such as magnetic memories or flux qubits.
\end{abstract}

\date{June 15, 2021}%

\maketitle

\section{Introduction}
Superconducting nanobridges with large sheet resistances in the normal state can provide a large kinetic inductance $L_K$ \cite{Tin96,Ann10}. The large kinetic inductance results from the kinetic energy of the supercurrent charge carriers and, in contrast to the geometric self-inductance, it does not couple to a magnetic field \cite{Cau16}. Moreover, the kinetic inductance is nonlinear in both current and temperature. These unique properties of high kinetic inductance nanobridges and nanowires result in their application as scalable key elements in many recently demonstrated device applications ranging from single-photon detectors \cite{Day03} to qubit readout and qubit architectures \cite{Moo99,Man09,Eom12,Pel18,Sch20}, magnetic memories and sensors \cite{Luo14,Mur17} and superconductor microwave detectors \cite{Zmu12,Coi20}. Despite the technical relevance and many applications of high kinetic inductance devices, it is complicated to precisely measure the kinetic inductance value. The existing methods to extract this value either require complex device structures like resonator circuits or rely on the total inductance's temperature dependence to separate geometric and kinetic contributions \cite{Mes69,Shi04,Wan18,Cai20}.

In this work, we show a straightforward way to determine the kinetic inductance of a nanobridge. Using this method, we conduct an experimental study of lithographically fabricated Mo$_{79}$Ge$_{21}$ superconducting nanobridges and their current phase relation (C$\Phi$R). The latter is approximately given by \cite{Lik79,Has02,Haz13,Mur17b,Haz19}

\begin{equation} \label{eq:1}
    I_s=\frac{\mathrm{\Phi}_0}{2\pi}\frac{1}{L_K}\ \varphi.
\end{equation}
Here, $I_s$ is the supercurrent through the nanobridge, $L_K$ is its kinetic inductance and $\varphi$ is the phase difference across the nanobridge, which is limited by a critical value $\varphi_c$ above which the bridge transits to the normal state. We do this by imbedding the nanobridges in a SQUID \cite{Roc07,Bur14,Mur17b,Col21}. The response of the SQUIDs used in this work is completely determined by their kinetic inductance, making the critical current versus magnetic field oscillations $I_c(B)$ of the SQUIDs directly reflect the C$\Phi$R and hence also the kinetic inductance of the nanobridge \cite{Mur17b,Col21}. 

We observe that for $T \ll T_c$, the C$\Phi$R is linear everywhere apart from the region where both SQUID arms near their critical phase difference. This nonlinearity can be captured by introducing a nonlinear kinetic inductance in equation \ref{eq:1}, quadratic in the current and originating from kinetic suppression of the condensate. We show that for devices of the same thickness, the $L_K$ values scale with nanobridge dimensions as $ \sim L/W$ for the lithographically fabricated nanobridges. Here $L$ and $W$ are the length and width of the nanobridge, respectively. Furthermore, we demonstrate that the SQUID $I_c(B)$ characteristic is tuneable through lithographic control over the nanobridge dimensions. In this way, SQUID properties like the SQUID sensitivity and the positions of the critical current maxima can be controlled. These observations are beneficial for future superconducting device design and operation (e.g. magnetic memories and qubit readout).

\section{Nanobridge SQUIDs}
\begin{figure}[!htb]
        \center{\includegraphics[width=0.5\textwidth]
        {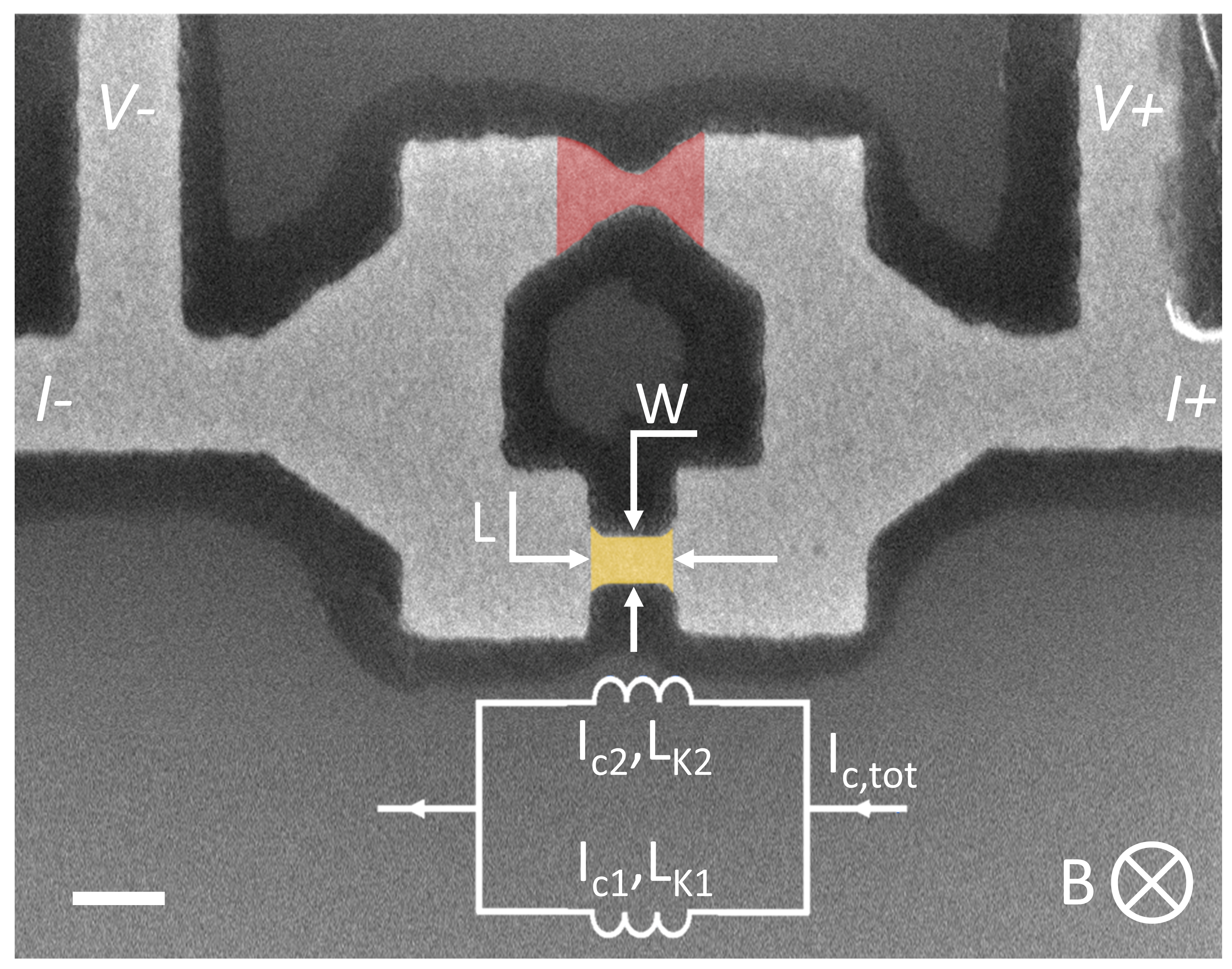}}
        \caption{ Scanning electron microscopy (SEM) image of a prototypical SQUID device. The area indicated in red corresponds with the Dayem bridge, while the yellow area indicates the nanobridge. The latter’s width $W$ and length $L$ are indicated. The white scale bar represents 200 nm. The four-point current and voltage contacts are indicated as $I\pm$ and $V\pm$. The white circuit diagram presents an equivalent electronic circuit of the SQUID. $L_{K1}$ and $L_{K2}$ represent the inductances of each branch, while $I_{c1}$ and $I_{c2}$ represent the two critical currents of each branch. $I_{c,tot}$ is the total critical current of the SQUID. The applied magnetic field $B$ is oriented as shown in the figure.}
        \label{fig:SEM}
\end{figure}

Figure \ref{fig:SEM} shows a scanning electron microscopy image of a prototypical nanobridge SQUID device, which serves as a platform to study the dependence of the nanobridge C$\Phi$R on the bridge dimensions. The fabricated structures are: (i) SQUIDs containing two Dayem bridges (see inset of figure \ref{fig:dev_g}) and (ii) SQUIDs containing one Dayem bridge (indicated in red in figure \ref{fig:SEM}) and one well-defined nanobridge weak link (indicated in yellow in figure \ref{fig:SEM}) of which the dimensions (length $L$ and width $W$) are varied. The sample geometry is defined using conventional electron beam lithography, followed by pulsed laser deposition of a 25 nm/5 nm thick MoGe/Au film and a standard lift-off process \cite{Tim16,Pan19}. The gold layer protects the devices from oxidation. The devices all have a similar superconducting to normal state transition temperature of approximately $T_c\sim $ 6 K. From measurements of the superconducting to normal state phase boundary on similarly prepared plain films of MoGe/Au, the coherence length can be determined and is approximately given by $\xi$($T$ = 0 K) $\sim$ 10 nm \cite{Mot14}.

For nanobridges in the dirty limit and for $T \ll T_c$, the kinetic inductance $L_{K}$ is proportional to $\hbar R_\blacksquare / (k_BT_c)$, where $R_\blacksquare$ is the sheet resistance \cite{Tin96,Ann10}. This implies that high kinetic inductances can be achieved using materials with high sheet resistances. Similar to disordered superconductors  such as NbTiN ($\rho$ = 170 $\mu\Omega$ cm) \cite{Haza19}, TiN ($\rho$ = 100 $\mu\Omega$ cm) \cite{Pel18,She18} NbN \cite{Nie19} and granular Al \cite{Rot17}, the MoGe structures discussed in this work have a high estimated resistivity ($\rho$ = 219 $\mu\Omega$ cm), leading to a high kinetic inductance \cite{Tas08}. Another advantage of using MoGe is the fact that it can be fabricated as a thin homogenous amorphous film \cite{Bez00}. For the kinetic inductance to dominate, the loop size should be small so as to minimize the geometrical contribution to the inductance. For the loop size used, the geometric inductance can be estimated as $\approx$ 2 pH \cite{Gre74}.

\section{Properties of a SQUID containing two Dayem bridges} \label{sec:twoDayem}
\begin{figure}[!htb]
        \center{\includegraphics[width=0.5\textwidth]
        {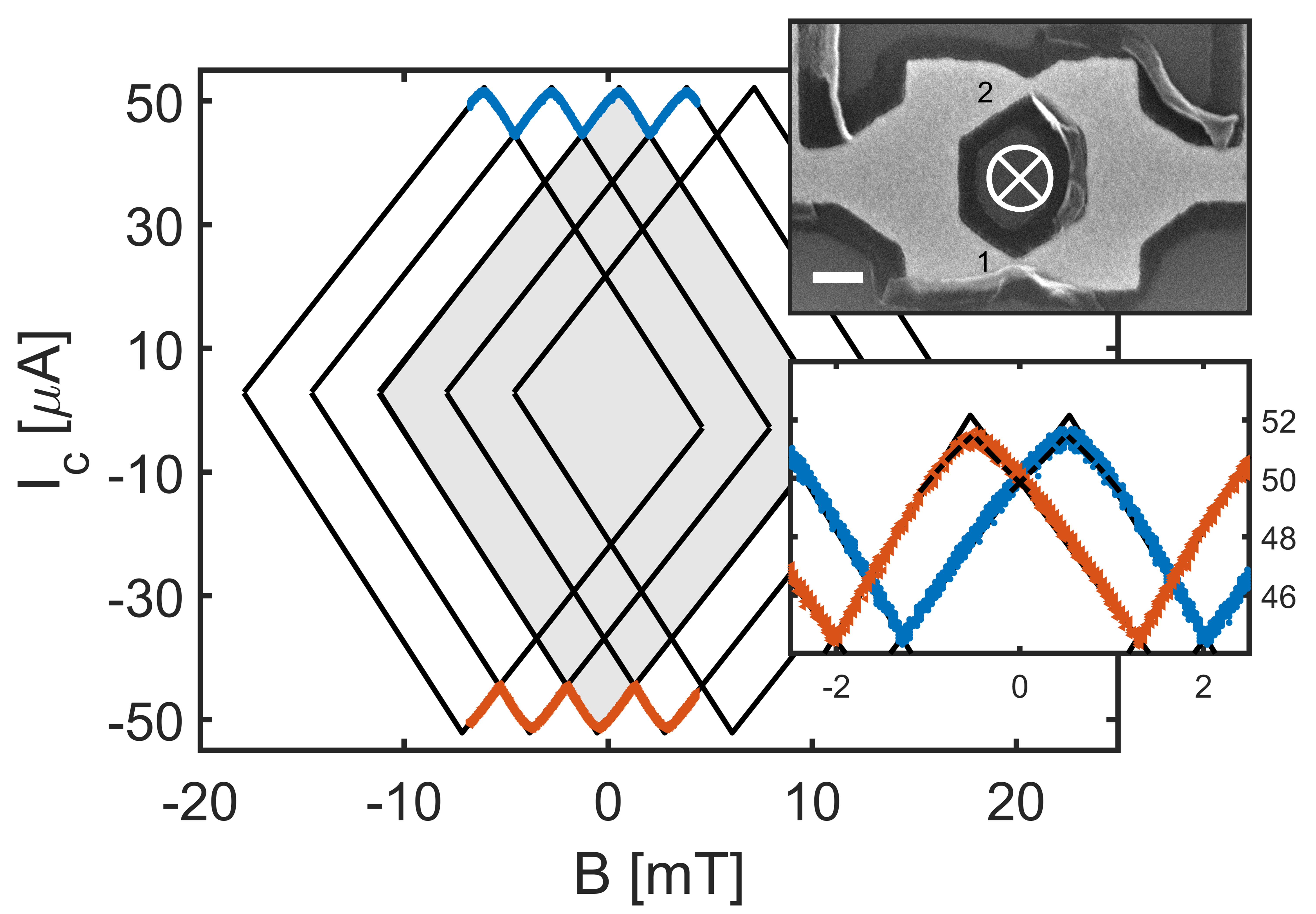}}
    \caption{The critical currents against magnetic field for device g, which contains two Dayem bridges. Measured critical currents for positive (negative) bias are shown in blue (red). The solid lines represent the vorticity diamonds generated by the model (see text). The fitting parameters of the vorticity diamonds are $I_{c1}$ = 27.5 $\pm$ 0.8 $\mu$A, $\varphi_{c1}$ = 10.2 $\pm$ 0.1 rad, $I_{c2}$ = 24.7 $\pm$ 0.9 $\mu$A, $\varphi_{c2}$ = 11.2 $\pm$ 0.1 rad, $L_{K1}$ = 122 $\pm$ 4 pH and $L_{K2}$ = 149 $\pm$ 6 pH. The $n_v=0$ vorticity diamond is indicated in grey. The bottom inset shows a zoom of the top (bottom) vertices of the diamonds - the positive currents are again shown in blue, the absolute values of the negative critical currents in red. The dotted lines show how taking a nonlinear $L_K$ into account can capture the shape of the diamond top. The top inset shows a scanning electron microscopy image of the investigated device, the scale bar corresponds with 200 nm. The applied magnetic field $B$ is oriented as shown in the figure. The bottom SQUID arm corresponds with $j$ = 1 in equation \ref{eq:2}, the top SQUID arm with $j$ = 2.}
    \label{fig:dev_g}
\end{figure}

First, the properties of a SQUID containing two Dayem bridges are examined. This allows to obtain the properties of the Dayem bridges used in this work \cite{And64}, which will act as reference junctions to study the properties of nanobridge type weak links. A representative scanning electron microscopy image is shown in the top inset of figure \ref{fig:dev_g}. Several SQUID devices of this type were fabricated and characterized. All measured devices exhibit comparable normal state resistances from which the resistance of a single Dayem bridge arm can be extracted as $R_{Day,arm}$ = 703 $\pm$ 19 $\mathrm{\Omega}$. Despite the small resistance spread of 3$\%$, it is clear from scanning electron microscopy imaging and from the spread of the measured device critical currents that the fabrication process results in an unavoidable spread of the Dayem bridge junction parameters. 

Figure \ref{fig:dev_g} shows the measured critical current versus field data $I_c(B)$ of a prototypical device, device g. The red branches $I_c^-(B)$ indicate the field dependence of the critical current obtained when sweeping the current from zero (the superconducting state) towards a large negative bias current that drives the device into the normal state, whereas the blue branches $I_c^+(B)$ are obtained when sweeping the current from zero towards a large positive bias current. All measurements were performed at 300 mK. At each magnetic field value, a set of 100 $VI$ measurements in both sweep directions is obtained using a current ramp rate of 3.6 mA/s. For each of these curves the critical current was then extracted by means of a voltage criterium of 0.5 mV, which is taken just above the noise level. As the transition is very sharp, the obtained critical currents does not depend on the chosen voltage criterium (at least below 2 mV). The critical currents extracted in this way are shown in figure \ref{fig:dev_g} as small dots. Due to thermal or quantum fluctuations, phase slip events cause a premature escape from the superconducting state before the depairing current is reached, resulting in a stochastic distribution of the critical current around an average value \cite{Sah09,Are12}. 

The shape of the oscillations can be captured by the vorticity diamond model introduced in Ref. \cite{Mur17b}. This model describes a SQUID containing two weak links, which both have a linear C$\Phi$R

\begin{equation}
    I_j\ =\ I_{cj}\frac{\varphi_j}{\varphi_{cj}},
    \label{eq:2}
\end{equation}
where $I_j$ represents the supercurrent through the $j$-th weak link, with $j$ = {1,2}, and where $\varphi_j$ is the phase difference of the macroscopic wavefunction taken between the end points of the $j$-th weak link. Further, $I_{cj}\geq0$ is the critical current and $\varphi_{cj}\geq0$ is the critical phase difference at which the weak link switches to the dissipative state. 

From equation \ref{eq:2} and the second Josephson relation \cite{Jos62,Jos65}, it is clear that the $j$-th weak link behaves as an inductor with kinetic inductance $L_{Kj}=\frac{\mathrm{\Phi}_0}{2\pi}\frac{\varphi_{cj}}{I_{cj}}$, where $\mathrm{\Phi}_0$ is the magnetic flux quantum (see equation \ref{eq:1} and the electrical model in figure \ref{fig:SEM}). For the typical lengths of the Dayem bridges explored in this work (approximately 150 nm), a linear C$\Phi$R represents a reasonable approximation in the explored low temperature range $T \ll T_c$ \cite{Haz19}. Indeed, by solving the Ginzburg-Landau equations, it has been shown that for nanobridges longer than 3$\xi(T)$, the C$\Phi$R becomes multivalued and progressively more linear \cite{Has02,Pod07}. Although the Ginzburg-Landau formalism is strictly valid only close to $T_c$, an almost linear C$\Phi$R has been predicted for thin and long wires even at $T$ = 0 K (see Ref. \cite{Mur17} and references therein). The diamond model does not consider any nonlinear dependences of the kinetic inductance on current or temperature and treats the average critical current and the ideal, fluctuation-free depairing current as if they are equal.

The total current through the SQUID is given by:
\begin{equation}\label{eq:3}
    I=I_{c1}\frac{\varphi_1}{\varphi_{c1}}+I_{c2}\frac{\varphi_2}{\varphi_{c2}}.
\end{equation}
As the superconducting order parameter must be single-valued, the total acquired phase difference around the superconducting loop must be an integer multiple of 2$\pi$. When applying an external magnetic field $B$ perpendicular to the SQUID loop (with an orientation as indicated in figure \ref{fig:SEM}), the phase differences across each wire and the electrodes must therefore add up as:

\begin{equation}\label{eq:4}
    \varphi_1-\varphi_2+2\pi\frac{B}{\mathrm{\Delta} B}=2\pi n_v .
\end{equation}
Here, $\mathrm{\Delta} B$ is the Little-Parks oscillation period \cite{Hop05} while $n_v$ is the vorticity or winding number of the loop. Note that in equation \ref{eq:4} it is assumed that any contribution from the geometric inductance of the SQUID to $B$ can be neglected, effectively decoupling equations \ref{eq:3} and \ref{eq:4}. Combining these two equations with the requirement that superconductivity should be destroyed if $\text{abs}(\varphi_j)>\varphi_{cj}$ in any of the bridges, one can calculate the total critical current of the SQUID for a given vorticity $n_v$ and applied magnetic field $B$. The total critical current of the SQUID $I_c(B,n_v)$ is assumed to be the smallest total applied current at which the current across either wire reaches its critical value. 

\begin{figure*}[!htb]
        \center{\includegraphics[width=\textwidth]
        {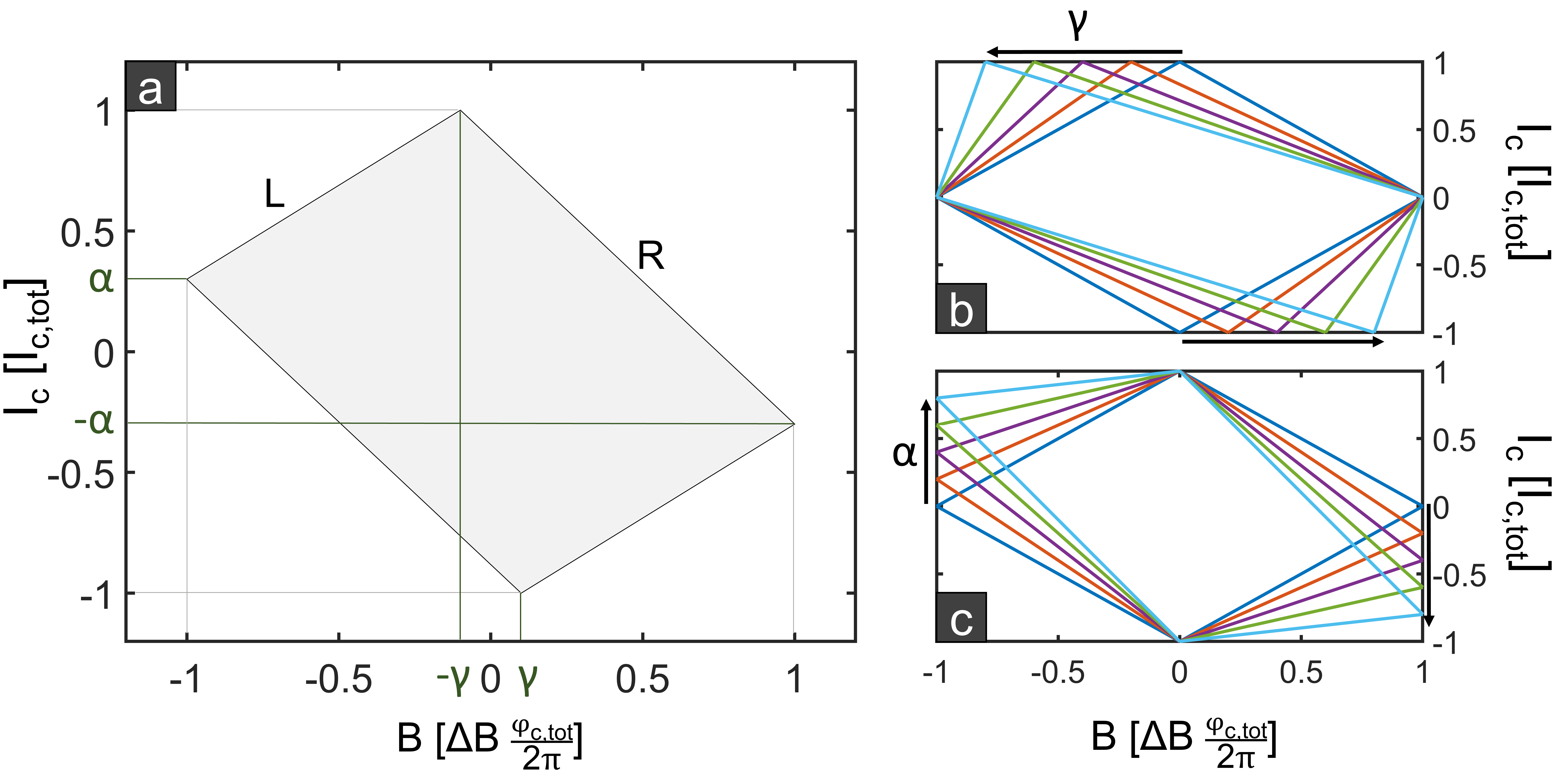}}
        \caption{(a) The $n_v=0$ vorticity diamond for a current asymmetry $\alpha=0.3$ and an phase asymmetry $\gamma=0.1$. The vertices of the diamonds are determined by the asymmetries in the critical currents and critical phase differences. The diamond extends over a range $B=\Delta B \varphi_{c,tot}/\pi$ in magnetic field. The $n_v$-th vorticity diamond is identical to the   $n_v=0$ diamond, but shifted along the magnetic field axis by $B=n_v\mathrm{\Delta} B$. (b) Evolution of the diamond shape for $\alpha=0$ and increasing $\gamma$  from 0 to 0.8 in steps of 0.2, as indicated by the arrows. (c) Evolution of the diamond shape for $\gamma=0$ and for increasing $\alpha$ from 0 to 0.8 in steps of 0.2, as indicated by the arrows.}
        \label{fig:model}
\end{figure*}

For each value of $n_v$, the solution for $I_c(B,n_v)$ forms a so-called vorticity diamond. As illustrated in figure \ref{fig:model}a, the magnetic field range and the range of critical currents in which the $n_v=0$ vorticity state exists (i.e. the vertices of the diamond) are determined by the asymmetry in critical currents $\alpha=\left({I}_{c1}-I_{c2}\right)/I_{c,tot}$, with $I_{c,tot}={I}_{c1}+I_{c2}$, and the asymmetry in critical phase differences $\gamma=\left(\varphi_{c1}-\varphi_{c2}\right)/\varphi_{c,tot}$, with $\varphi_{c,tot}=\varphi_{c1}+\varphi_{c2}$. The separate influences of varying phase and current asymmetries are shown in figure \ref{fig:model}b and figure \ref{fig:model}c, respectively. The $n_v$-th vorticity diamond is identical to the $n_v=0$ diamond shown in figure \ref{fig:model}a, but is shifted along the magnetic field axis by $B=n_v\mathrm{\Delta}B$. As the vorticity diamond extends over a range $B=\Delta B \varphi_{c,tot}/\pi$ in magnetic field, diamonds of adjacent vorticities overlap if $\varphi_{c,tot}>\pi$, (i.e. twice the critical phase difference of a conventional tunnel junction, $\varphi_{c,tunnel}=\pi/2$), resulting in a multivalued critical current.
An important property of the vorticity diamond is that on each branch of the vorticity diamond, the phase difference over one arm of the SQUID remains constant, while the phase difference over the other arm varies linearly with the applied magnetic field. Consequently, each branch of the $I_c \left(B\right)$ curve immediately reflects the C$\Phi$R of one arm of the SQUID \cite{Roc07}. For the branches $L$ and $R$ indicated in figure \ref{fig:model}a, the magnetic field dependence of the phase differences $\varphi_j$ over the weak links and the field dependence of the critical current $I_c(B)$ dependence are given by:

\begin{align} \label{eq:branches_behaviour}
   L: &\left\{
    \begin{array}{ll}
         \varphi_1 = \varphi_{c1}, \\
         {{ \varphi}_2(B)=\varphi}_{c1}+2\pi\frac{B}{\mathrm{\Delta} B}, \\
         I_c\left(B\right)=I_{c1}+\left(\frac{1}{L_{K2}}\frac{\Phi_0}{2\pi}\right)\varphi_{2},
    \end{array}
\right.\\
    R: &\left\{
    \begin{array}{ll}
         \varphi_2 = \varphi_{c2}, \\
         {{ \varphi}_1(B)=\varphi}_{c2}-2\pi\frac{B}{\mathrm{\Delta} B}, \\
         I_c\left(B\right)=I_{c2}+\left(\frac{1}{L_{K1}}\frac{\Phi_0}{2\pi}\right) \varphi_{1}.
    \end{array}
\right. \nonumber
\end{align}
As the $n_v=0$ vorticity diamond is point-symmetric around the origin, similar expressions exist for the corresponding opposite branches. Considering the field and current orientation with respect to the sample surface used in the experiment, the SQUID's transition to the normal state for branch $L(R)$ corresponds with arm $j$ = 1(2) reaching its critical current, corresponding with a critical phase difference $\varphi_1=\varphi_{c1} ({{\varphi}_2=\varphi}_{c2})$. As such, branch $L(R)$ reflects the C$\Phi$R of arm $j$ = 2(1). Equation \ref{eq:branches_behaviour} also shows that the kinetic inductances $L_{Kj}$ of the SQUID arm $j$ = 1(2) are inversely related to the diamond slope of branch $R(L)$.


In figure \ref{fig:dev_g} the solid black lines result from a fit of the $I_c(B)$ data to the model described above. It is clear that the model captures the $I_c(B)$ characteristic well. Note that the kinetic inductances are determined by the slope of the linear part of each branch of the vorticity diamond. A first observation is that for this symmetrically designed SQUID the asymmetries in the critical currents ($\alpha$ = 0.06), the critical phase differences ($\gamma$ = -0.05) and the kinetic inductances ($L_{K1}$ = 122 pH and $L_{K2}$ = 149 pH) are small. The kinetic inductances are much larger than the geometric contribution to the inductance ($\sim$ 2 pH) \cite{Gre74}, which is a key assumption in the used model. The critical phase differences of both Dayem bridges are far larger than $\pi/2$: $\varphi_{c1}$ = 10.2 rad and $\varphi_{c2}$ = 11.2 rad. This corresponds with the observation that their length exceeds 3$\xi\left(T\right)$. As introduced in the section above, the $I_c(B)$ dependence of each branch of the vorticity diamond is in one-to-one correspondence with the C$\Phi$R of the corresponding Dayem bridge. The majority of the experimental data is well-captured by the linear edges of the vorticity diamond, implying that the C$\Phi$R of the Dayem bridges is indeed approximately linear. However, close to the maximum critical current or maximum critical phase difference (i.e. the top and bottom vertices of the vorticity diamond, see bottom inset of figure \ref{fig:dev_g}), the critical current deviates from a linear dependence on the applied magnetic field for both branches. This reflects that the C$\Phi$R is nonlinear in this current range. The nonlinear dependence can be described by introducing a nonlinear kinetic inductance and can be attributed to the kinetic suppression of the condensate due to the presence of a supercurrent \cite{Ant03}. For $T \ll\ T_c$, the kinetic inductance of a superconducting strip can be expanded in terms of the depairing current $I_{dep}(B)$ as
\begin{equation}
    L_{K}\left(I_{dep}\right)= L_K\left(0\right)\left[1\ +\ \frac{I_{dep}^2}{I_\ast^2}\ +\ ...\right],
\end{equation}
where $I_\ast$ is a constant which sets the scale of the quadratic nonlinearity \cite{Zmu12}. No odd-ordered terms appear due to symmetry considerations, as the bridge must have the same kinetic inductance regardless of the current orientation. Within the Usadel framework, Ref. \cite{Ant03} provides an estimate for the magnitude of the quadratic nonlinearity as $I_\ast=4.7I_{dep}/\sqrt{1.9}$, indicating that the nonlinearity becomes important only at currents close to the depairing current of the strip. In Ref. \cite{Luo14}, this nonlinearity of the kinetic inductance was exploited to fabricate an ultrasensitive magnetometer.

To summarize, the vorticity diamond's branches directly reflect the bridge C$\Phi$Rs, the vertices are determined by the asymmetries in critical currents $\alpha$ and critical phases $\gamma$ and the nonlinearity at the top vertex is understood to be caused by depletion of the superconducting condensate. In the next section, this approach is used to explore the impact of geometry on the C$\Phi$R of the fabricated nanobridges.

\section{Properties of SQUIDs containing one Dayem bridge and one nanobridge}
\begin{figure}[!htb]
        \center{\includegraphics[width=0.5\textwidth]
        {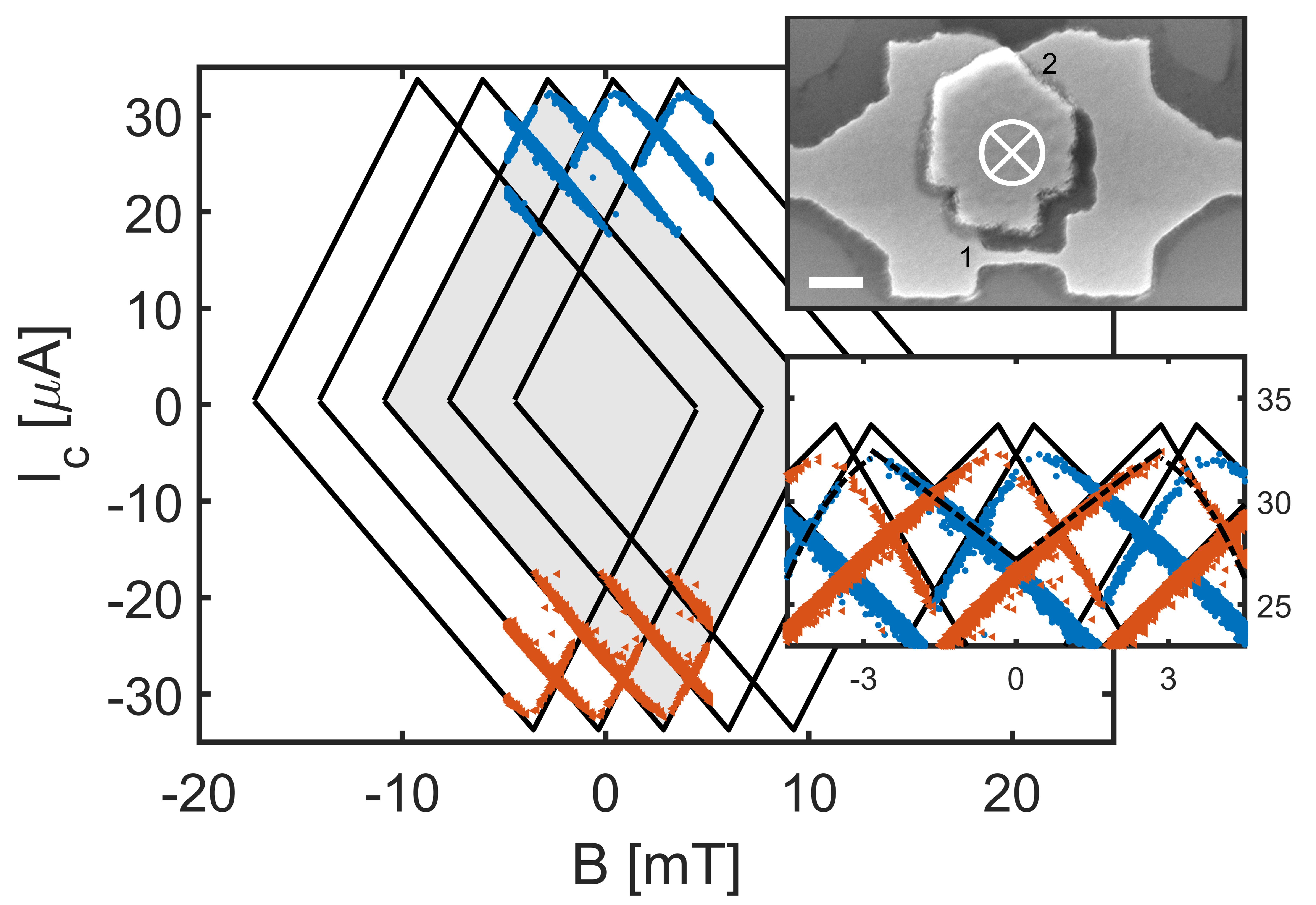}}
    \caption{The critical currents against magnetic field for device u, which contains a nanobridge of length $L \sim$ 319 nm and width $W \sim$ 34 nm. Measured critical currents for positive (negative) bias are shown in blue (red). The solid lines represent the vorticity diamonds generated by the model. The fitting parameters of the vorticity diamonds are $I_{c1}$ = 17.1 $\pm$ 0.8 $\mu$A, $\varphi_{c1}$ = 13.5 $\pm$ 0.6 rad, $I_{c2}$ = 16.7 $\pm$ 0.6 $\mu$A, $\varphi_{c2}$ = 7.9 $\pm$ 0.6 rad, $L_{K1}$ = 261 $\pm$ 17 pH and $L_{K2}$ = 156 $\pm$ 15 pH. The $n_v=0$ vorticity diamond is indicated in grey. The bottom inset shows a zoom of the top (bottom) vertices of the diamonds - the positive currents are again shown in blue, the absolute values of the negative critical currents in red. The dotted lines show how taking a nonlinear $L_K$ into account can capture the shape of the diamond top. The top inset shows a scanning electron microscopy image of the investigated device, the scale bar corresponds with 200 nm. The applied magnetic field $B$ is oriented as shown in the figure. The bottom SQUID arm corresponds with $j$ = 1 in equation \ref{eq:2}, the top SQUID arm with $j$ = 2. Due to incomplete lift-off, the inner part of the SQUID is not completely removed. }
    \label{fig:dev_u}
\end{figure}

This section examines the properties of fabricated SQUIDs containing one Dayem bridge and one nanobridge (see figure \ref{fig:SEM}), upon changing the nanobridge dimensions $(L,W)$. The different device dimensions as measured by scanning electron beam microscopy are shown in table \ref{tab:comparison}. The total normal state resistance of each device, together with the average value of the resistance of a SQUID arm containing a Dayem bridge, allows to determine the resistance of the SQUID arm containing the nanobridge for each device. These resistance values are consistent with the nanobridge geometries. Figure \ref{fig:dev_u} shows the critical current versus field data $I_c(B)$ for a particular SQUID device (device u), which contains one Dayem bridge and one long nanobridge ($L\sim\ $ 319 nm). The measurement method is the same as for device g, with a current ramp rate of 3.5\ mA/s. For the typical lengths of the nanobridges and temperatures explored in this work ($L \sim $ 100-340 nm $\gg$ 3$\xi$, $T$ = 300 mK $\ll$ $T_c$), it is expected that the C$\Phi$R of the nanobridge is quasi-linear \cite{Has02,Pod07,Mur17}. The resulting fit using the model described in section \ref{sec:twoDayem} is shown in solid black lines. It is clear that once again, the model captures the $I_c(B)$ characteristic well.

\begin{table*}
\begin{center}
    
    \begin{ruledtabular}
    \caption{The nanobridge dimensions (length $L$ and with $W$) are listed, together with the fit parameters of each device: each SQUID arm is decribed by its critical current $I_{cj}$, critical phase difference $\varphi_{cj}$ and the linear part of its inductance $L_{Kj}$. Device g consists of two Dayem bridges and so no nanobridge dimensions are given.}
    \label{tab:comparison}
\begin{tabular}{c c c c c c c c c} 
    device & L [nm] & W [nm] & $I_{c1}$ [$\mu$A] & $I_{c2}$ [$\mu$A] & $\varphi_{c1}$ [rad]   & $\varphi_{c2}$ [rad] & $L_{K1}$ [pH] & $L_{K2}$ [pH]  \\ \hline
g & &  &27.5 $\pm$ 0.8 & 24.7 $\pm$ 0.9 & 10.2 $\pm$ 0.8  & 11.2 $\pm$ 0.1 &  122 $\pm$ 4 & 149 $\pm$ 6 \\ 
c  & $101 \pm 5$&  58 $\pm$ 5 & 58 $\pm$ 3 & 39 $\pm$ 4 & 15 $\pm$ 2  & 12 $\pm$ 2 & 84 $\pm$ 12 & 105 $\pm$ 21 \\
m   &$176 \pm 5$& 54 $\pm$ 5 &  49.9 $\pm$ 0.9 & 39  $\pm$ 1 &  18.4 $\pm$ 0.2 & 15.4 $\pm$ 0.2 & 121 $\pm$ 3 & 131 $\pm$ 4 \\
u   &$319 \pm 5 $& 34 $\pm$ 5 & 17.1 $\pm$ 0.8 & 16.7 $\pm$ 0.9 & 13.5 $\pm$ 0.6 & 7.9 $\pm$ 0.6 & 261 $\pm$ 17 & 156 $\pm$ 15 \\ 
v   &$313 \pm 5$& 45 $\pm$ 5 & 23.3 $\pm$ 0.8 & 23 $\pm$ 1 & 15.2 $\pm$ 0.7 & 10.9 $\pm$ 0.7 & 215 $\pm$ 12 & 158 $\pm$ 14 \\ 
w   &$338 \pm 5$& 38 $\pm$ 5 & 28 $\pm$  1 & 29 $\pm$ 1 & 19 $\pm$ 1  & 13 $\pm$ 1 & 225 $\pm$ 14 & 147 $\pm$ 13 \\ 
\end{tabular}
\end{ruledtabular}
    
\end{center}
\end{table*}

As introduced in section \ref{sec:twoDayem}, the $I_c(B)$ dependence of each branch of the vorticity diamond is in one-to-one correspondence with the C$\Phi$R of the corresponding Dayem or nanobridge. Similarly as for the Dayem bridges in section \ref{sec:twoDayem}, the experimental data is well-captured by the linear edges of the vorticity diamond. This reflects that the C$\Phi$R of the nanobridge is approximately linear. Only close to the maximum critical current or maximum critical phase difference (i.e. the top vertices of the vorticity diamonds, see inset in figure \ref{fig:dev_u}), the C$\Phi$R is nonlinear. This can be captured by introducing a nonlinear inductance into equation \ref{eq:branches_behaviour} (see dotted line in the inset of figure \ref{fig:dev_u}). The kinetic inductances obtained from the fitting ($L_{K1}$ = 261 pH and $L_{K2}$ = 156 pH) are substantially larger than the geometric contribution to the inductance ($\sim$ 2 pH) \cite{Gre74}. Since the kinetic inductance scales with the length and width of the bridge as $L_K \sim L/W$, the kinetic inductance of the nanobridge $L_{K1}$ is about 65$\%$ larger than the kinetic inductance of the Dayem bridge, $L_{K2}$. This difference is reflected in the different slopes ($\sim\ 1/L_K$) of the branches of the vorticity diamond. The critical phase differences of both the Dayem bridge ($L\sim$ 150 nm) and nanobridge are far larger than $\pi/2$. The critical phase difference for a nanowire can be estimated to be \cite{Mur17},
\begin{equation} \label{eq:7}
   \varphi_{c,est}\sim (\pi/2) (L/2\xi)
\end{equation}
and scales with the length of the bridge. This is in correspondence with the observation that the critical phase difference of the nanobridge ($\varphi_{c1}$ = 13.5 rad) is about 70$\%$ larger than the one of the Dayem bridge ($\varphi_{c2}$ = 7.9 rad), which has a smaller effective length. However, for the nanobridge of this particular device u, which has a length $L\sim$ 319 nm, this estimate provides $\varphi_{c,est}\sim$ 25 rad, exceeding the value obtained from the fitting procedure. Note that this estimate is an overestimate as premature switching due to thermal and quantum fluctuations does not allow to reach the maximum fluctuation-free depairing current and corresponding critical phase difference \cite{Moo05} (see section \ref{sec:sweeprate}). For device u, the depairing current can be estimated as $I_{dep}=\frac{2.6k_BT_c}{e R_\xi}$ = 52 $\mu$A, where $R_\xi$ is the resistance of a length of wire equal to $\xi$ \cite{Moo05}. This value exceeds the experimentally measured critical current. The device incorporates both a Dayem bridge and nanobridge, which have different C$\Phi$Rs. This results in different kinetic inductances (slopes of the diamonds). The asymmetry in critical currents ($\alpha$ = 0.01) remains small, but there is a strong asymmetry in critical phase differences ($\gamma$ = 0.26). The latter relocates the top and bottom vertices of the vorticity diamond (see figures \ref{fig:model}b and \ref{fig:model}c). More specifically, the top of the $n_v=0$ diamond (indicated in grey) is clearly shifted from the $B$ = 0 mT axis towards negative field values. 

\begin{figure}[!htb]
        \center{\includegraphics[width=0.5\textwidth]
        {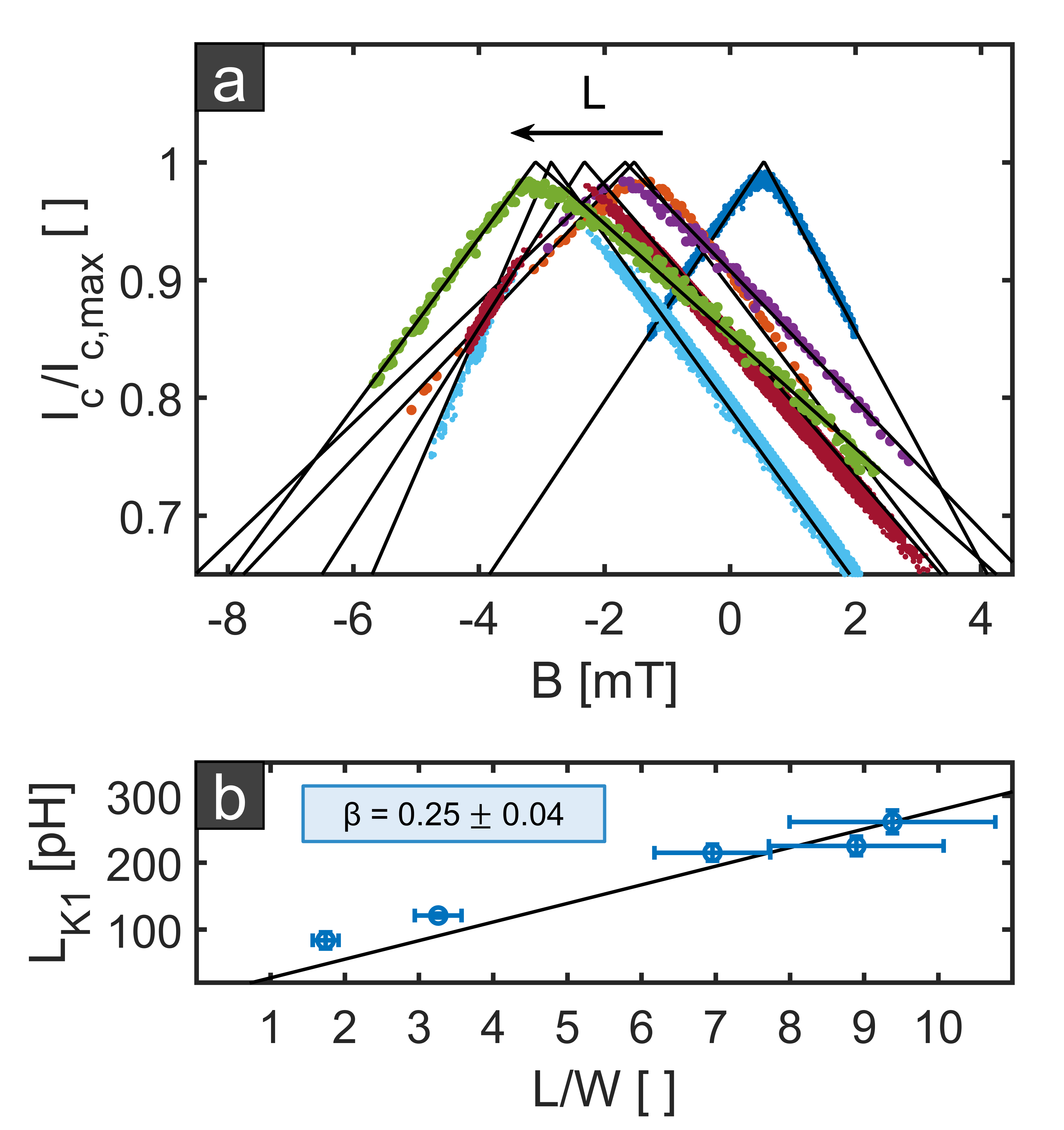}}
    \caption{(a) Field-dependence of the current-normalized $n_v=0$ vorticity diamond for device g (blue), c (red), m (purple), u (light blue), v (bordeaux) and w (green). The arrow on top shows the increase of the nanobridge length $L$. (b)  Kinetic inductance of the nanobridges as obtained from the fitting procedure versus aspect ratio of the bridge for devices c, m, u, v and w. The solid line shows a linear fit, using $L_K=\beta \frac{L}{W} \frac{{\hbar R}_\blacksquare}{k_BT_c}$, where $T_c$ = 6 K and $R_\blacksquare$ = 88 $\mathrm{\Omega}$. The fit yields $\beta$ = 0.25 $\pm$ 0.04.}
    \label{fig:comparison}
\end{figure}

Several SQUIDs containing one Dayem bridge and one nanobridge (see figure \ref{fig:SEM}) were investigated upon changing the dimensions ($L,W$) of the nanobridge. Figure \ref{fig:comparison}a shows the experimental data corresponding to the $n_v=0$ vorticity diamond for the different devices and the fit using the model in section \ref{sec:twoDayem}. In order to compare different devices, critical currents are normalized to the top of their fitted vorticity diamonds, $I_c^\ast(B)=\ I_c(B)/I_{c,max}$. Note that in order to obtain critical current values in a particular field range, data obtained using different measurement protocols were merged (see supplementary information) \cite{Mur17}. The estimated dimensions of the nanobridge from scanning electron microscopy and the obtained fit parameters using the model of section \ref{sec:twoDayem} are shown in table \ref{tab:comparison}.

It is clear that the slopes of the vorticity diamonds change upon changing the nanobridge dimensions. In figure \ref{fig:comparison}b, the dependence of the nanobridge kinetic inductance on the nanobridge aspect ratio $L/W$ is shown, together with a fit using the theoretical expectation for kinetic inductance of a nanobridge \cite{Zmu12}
\begin{equation}
    L_K= \beta \frac{L}{W} \frac{\hbar R_\blacksquare}{k_B T_c}.
\end{equation}
Here $T_c$ is approximately 6 K and the sheet resistance $R_\blacksquare$ is estimated to be 88 $\mathrm{\Omega}$. From the fit one obtains $\beta$ = 0.25 $\pm$ 0.04, whereas within the Ginzburg-Landau framework, theory suggest $\beta \sim$ 0.14-0.18 \cite{Moo05,Zmu12}. This indicates that the nanobridge kinetic inductance, and hence the diamond slope (so SQUID sensitivity), can be controlled by means of the bridge dimensions. In comparison to the lift-off procedure used to fabricate the samples in this work, a fabrication process which relies on etching could be beneficial to reduce the variation of device characteristics. Note that the sheet resistance is inversely proportional to the device thickness. The devices studied in this work have a relatively large thickness, but it is possible to grow homogenous MoGe films as thin as 1 nm \cite{Gra84,Lau01}, thereby providing an easy way to increase the kinetic inductance even more.

Moreover, the position of the maximum critical current in field shifts to more negative field values compared to device g, which corresponds to a larger difference in critcal phase differences $\varphi_{c1}-\varphi_{c2}$. In particular, it shifts to the left for increasing nanobridge lengths (see figure \ref{fig:comparison}a and equation \ref{eq:7}). It is not straightforward to quantitatively compare the critical phase differences and currents obtained from the fitting procedure (see table \ref{tab:comparison}), as some of the critical currents are obtained using a different current sweep rate. As thoroughly discussed in Ref. \cite{Bez10}, high current sweep rates have an effect on the maximum attainable critical current and corresponding critical phase difference. This effect is negligible for the sweep rates used in this section, as demonstrated in section \ref{sec:sweeprate}.

\section{Influence of sweep rate} \label{sec:sweeprate}
\begin{figure*}[!htb]
        \center{\includegraphics[width=\textwidth]
        {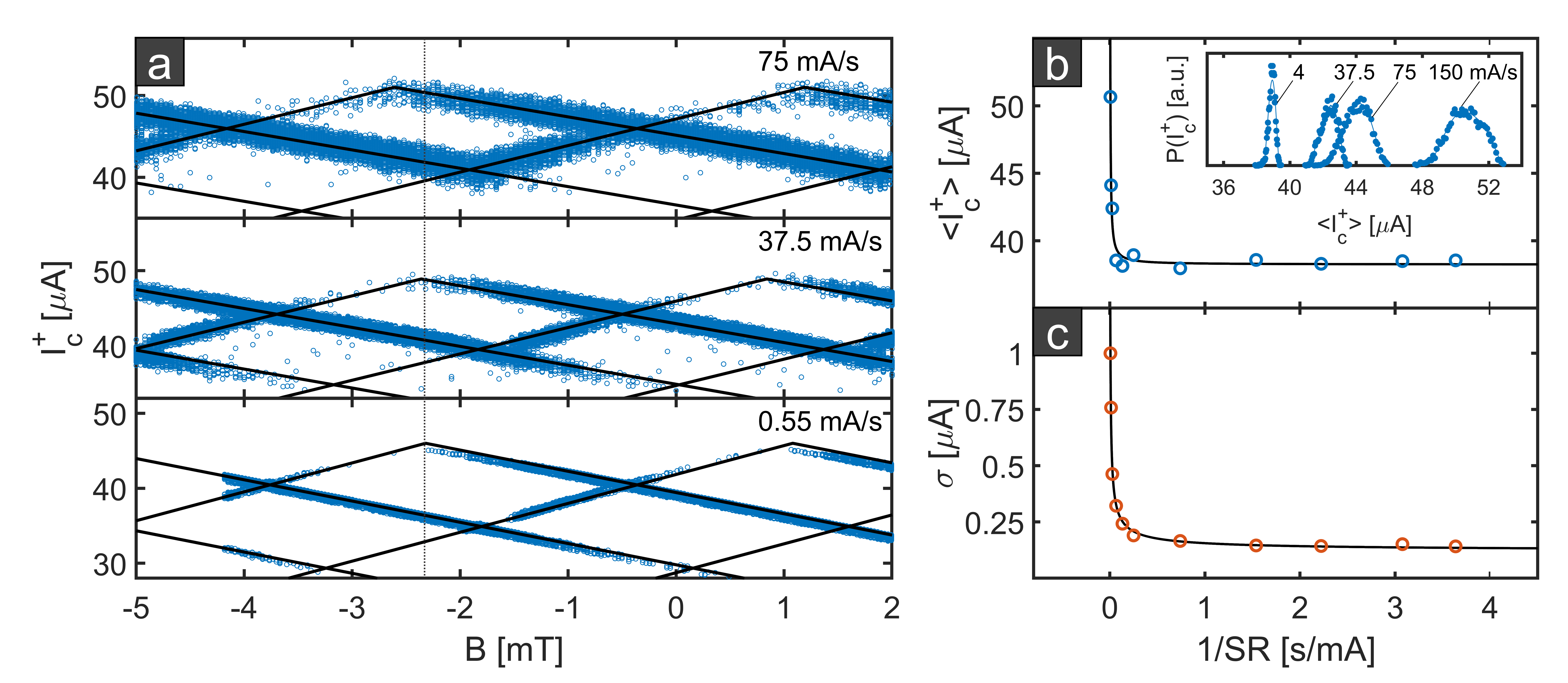}}
        \caption{(a) $I_c^+(B)$ characteristics of device v, shown for different sweep rates (0.55 mA/s, 37.5 mA/s and 75 mA/s). The grey dotted line shows that the current maximum position is only impacted by high sweep rates. (b) Dependence of the average nanobridge positive critical current on sweep rate ($SR$) for device v at $B$ = 0 mT, with sweep rates ranging from 0.275 mA/s to 150 mA/s. The inset shows the stochastic distribution for selected sweep rated values. (c) Dependence of the nanobridge positive critical current's standard deviation on sweep rate for device v at $B$ = 0 mT, with sweep rates ranging from 0.275 mA/s to 150 mA/s. At least 2500 critical currents per field value were collected in order to calculate the average critical current and its standard deviation. Above this value, both the average and the standard deviation converge towards an asymptotic value for all sweep rates.}
        \label{fig:sweeprate}
\end{figure*}

Figure \ref{fig:sweeprate}a shows the critical current $I_c^+(B)$ of one specific device (device v) for three different sweep rates (0.55 mA/s, 37.5 mA/s and 75 mA/s). The right side of the figure shows the average value $<I^+_c>$ (figure \ref{fig:sweeprate}b) and standard deviation $\sigma$ (figure \ref{fig:sweeprate}c) of the critical current at one specific field value ($B$ =\ 0\ mT), which corresponds to probing the nanobridge. The inset of figure \ref{fig:sweeprate}b shows the stochastic distribution for selected sweep rated values (4 mA/s, 37.5 mA/s, 75 mA/s and 150 mA/s).

It is clear that the maximum attainable critical current increases for the highest sweep rates. This can be explained as follows: due to thermal and quantum fluctuations of the superconducting phase (called phase slips), the sample transits to the normal state before reaching the maximum pair breaking current \cite{Sah09,Bau17}. In the ideal, fluctuation-free case, $\varphi_c$ in equation \ref{eq:1} is determined by the condition where $I_s$ reaches the depairing current limit. As such, the measured C$\Phi$R, which is sensitive to fluctuations, only reflects part of the ideal fluctuation-free C$\Phi$R. If one could sweep infinitely fast, the timescales of these fluctuations would be too large to have effect and the critical current would be equal to the depairing current. This means that by using higher sweep rates it is possible to better approach the ideal C$\Phi$R \cite{Bez10}. For this device, the depairing current can be roughly estimated as $I_{dep}\approx$ 60 $\mu$A \cite{Moo05}, which agrees with the data shown in figure \ref{fig:sweeprate}b. Figure \ref{fig:sweeprate}c illustrates that the standard deviation of the critical current distribution also increases for increasing sweep rates. This could indicate that fewer consecutive phase slip events are needed to switch to the normal state for high sweep rates (i.e. higher average critical current), which is directly linked to the stronger thermal footprint of a phase slip event at higher currents (i.e. higher average critical current) \cite{Tin03,Sha08,Pek09,Bau17}.



Figure \ref{fig:sweeprate}a shows that the shift of the vorticity diamond top is pronounced at higher sweep rates only and very small between 0.55 mA/s and 37.5 mA/s. The data shown in figure \ref{fig:comparison}a were obtained with sweep rates of maximally 4 mA/s, that is: in a regime where the influence of the sweep rate is negligible, making quantitative comparison between the different devices possible. Note that the diamond slope and thus also the kinetic inductance is insensitive to the current sweep rate.

\section{Conclusion}
In this work we studied MoGe superconducting nanobridges and their C$\Phi$R by imbedding them in a SQUID. The response of these SQUIDs is completely determined by their high kinetic inductance $L_K$, making the critical current versus magnetic field oscillations $I_c(B)$ of the SQUIDs directly reflect the C$\Phi$R of the nanobridge. For $T \ll T_c$, the C$\Phi$R is linear everywhere apart from close to the critical phase difference. This nonlinearity can be understood as kinetic suppression of the condensate. We demonstrated that the SQUID $I_c(B)$ characteristic is tuneable through lithographic control over the nanobridge dimensions. $L_K$ scales linearly with the nanobridge's aspect ratio $L/W$. This tunability and magnitude of $L_K$, together with its nonlinearity make our MoGe nanobridges extremely suitable for many applications, ranging from magnetic memories \cite{Mur17} to microwave detectors \cite{Zmu12,Coi20}. Furthermore, the measured $L_K$ can be maximized further by limiting the device thickness. This opens up the possibility to use the MoGe nanobridges as phase slip centers for phase-slip flux qubits \cite{Moo05}.

\begin{acknowledgements}
This work has been supported by the Research Foundation - Flanders (FWO, Belgium), with grant number G0B5315N.
\end{acknowledgements}

\bibliographystyle{unsrt}
\bibliography{bibfile}

\newpage
\onecolumngrid
\appendix

\renewcommand{\thefigure}{S\arabic{figure}}
\setcounter{figure}{0}

\section*{Appendix: Supplementary material}
All measurements were performed at 300 mK in an Oxford Instruments Heliox VL $^3$He cryostat. During these measurements, a room temperature $\pi$-filter with a cut-off frequency of 1 MHz was consistently used. \\

Flux oscillations were measured using a standard four-probe technique, using different protocols. We obtained $I_c(B)$ data by performing a standard VI-measurement at a particular $B$ value. This method only reveals part of the $I_c(B)$ curves, sometimes resulting in only a few datapoints for one particular branch. This makes it complex to perform an accurate fitting. To solve this problem and in order to generate more datapoints along each branch, a preparation protocol based on that introduced in Ref. \cite{Mur17} was employed. In this procedure, the concept of a ‘unique vorticity diamond’ is exploited. In the unique vorticity diamond, shown as a grey shaded region in figure \ref{fig:preparing} below, there is only one stable vorticity state.
\begin{figure}[!htb]
        \center{\includegraphics[width=0.95\textwidth]
        {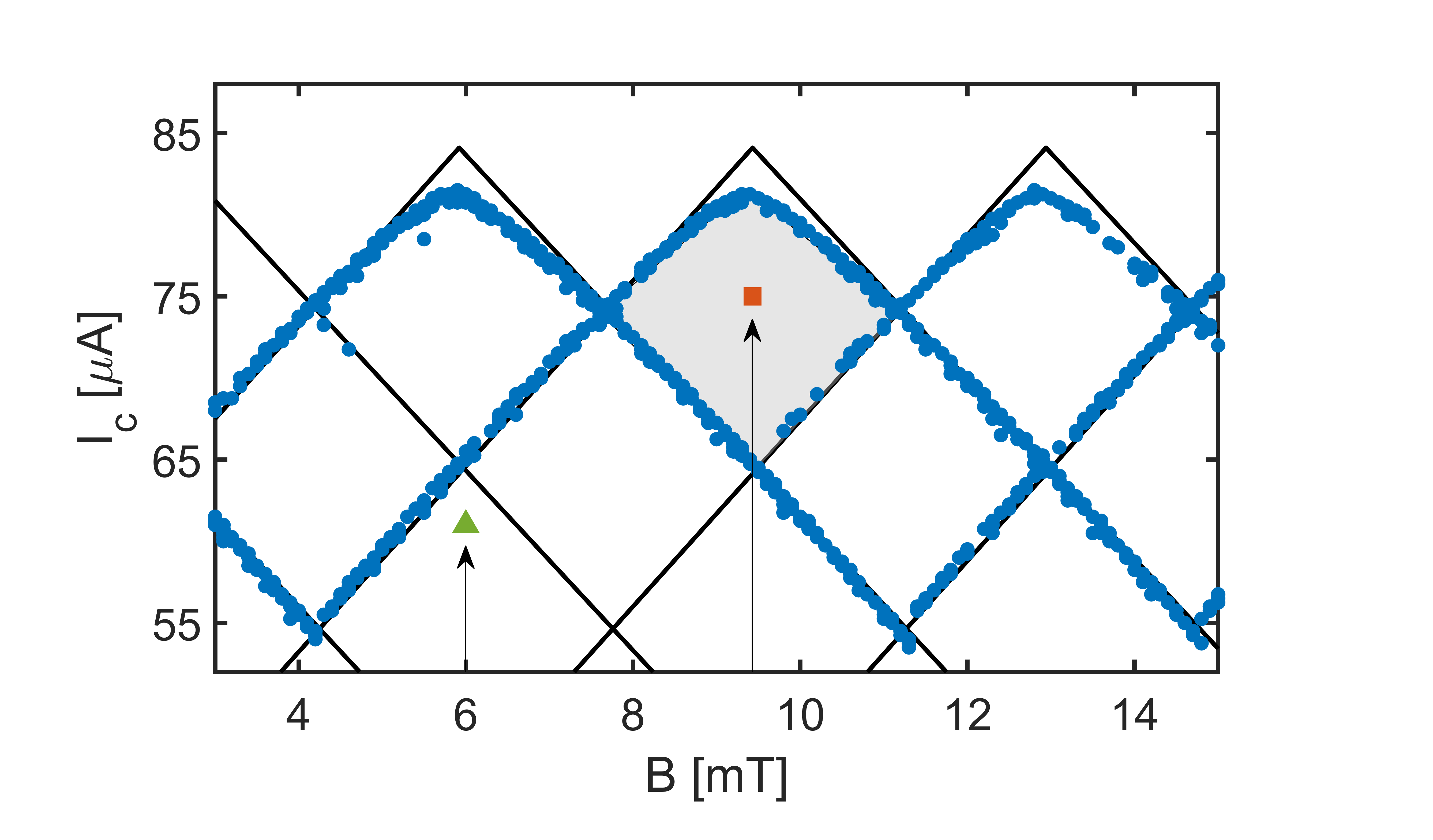}}
        \caption{Positive flux oscillations $I_c^+(B)$ of a MoGe nanobridge SQUID. The unique vorticity diamond of $n_v$ = 3 is indicated by the grey area. The system is prepared in the $n_v$ = 3 vorticity state by repeatedly applying a bias current in the unique vorticity diamond, as indicated by the red square. After that, the field is changed to the readout value under zero bias current. Once there, the critical current is measured by taking a regular VI, as shown by the green triangle.  By repeating this process, both branches of the diamond can be probed. The same procedure is applicable to the other vorticity diamonds.}
        \label{fig:preparing}
\end{figure}

To prepare the system in a specific vorticity state, we can apply an external magnetic field corresponding to the unique vorticity diamond field range. Then, we repeatedly apply a bias current which leads to switching to the normal states for all vorticity states except for the one associated with the unique vorticity diamond (shown by the red square). Once prepared, the system is driven to the readout field value under zero bias. When the readout value is reached, the critical is measured through a regular VI (indicated by the green triangle). This allows to probe a larger part of each vorticity diamond. \\

\begin{figure}[!htb]
        \center{\includegraphics[width=0.9\textwidth]
        {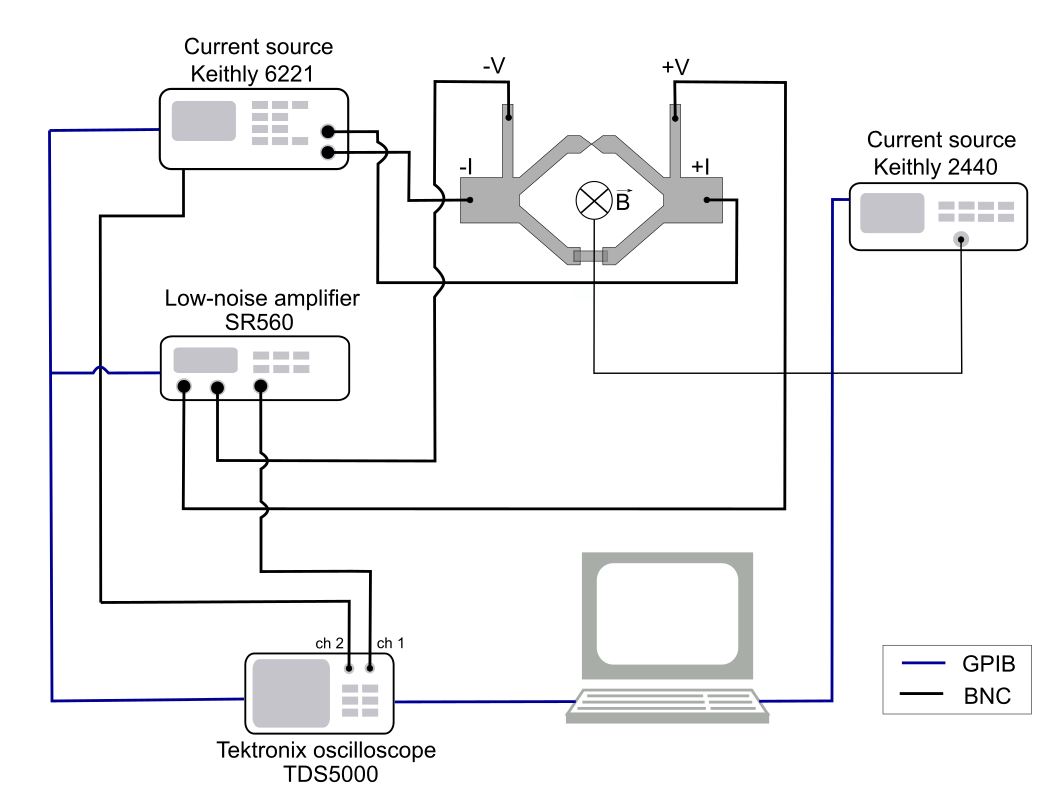}}
        \caption{Schematic overview of the AC-setup.}
        \label{fig:AC_setup}
\end{figure}

In order to generate more $I_c$ values at every field value, we used the setup shown in figure \ref{fig:AC_setup}. This setup allows to collect many VI-curves in a short period of time. For this, a triangular AC current wave signal was applied using a function generator (Keithley K6221). The current bias amplitude and frequency can be controlled, resulting in a controllable current sweep rate. The sample response was acquired via a low noise pre-amplifier (Stanford Research Systems SR560), after which the amplified signal passed on to a digital phosphor oscilloscope (Tektronix TDS5032B). LabView software was developed to automatically extract the switching current from the collected VI’s using a voltage criterium. The field was controlled through a current (supplied by a Keithely K2400) pushed through magnetic coils.

\end{document}